\documentclass[prd,showpacs,preprintnumbers,amsmath,amssymb,twocolumn,nofootinbib]{revtex4}
\usepackage[utf8]{inputenc}

\usepackage{graphicx}
\usepackage{color}
\usepackage{epstopdf}
\usepackage{hyperref}
\usepackage{cancel} 

\newcommand{\bea}{\begin{eqnarray}}
\newcommand{\eea}{\end{eqnarray}}

\newcommand{\eq}[1]{Eq.~\eqref{#1}}

\DeclareUnicodeCharacter{2212}{-}
\DeclareUnicodeCharacter{2217}{*}

\begin{document}
\preprint{PSI-PR-19-08, ZU-TH  25/19}

\title{Correlating Tauonic B Decays to the Neutron EDM via a Scalar Leptoquark}

\author{Andreas Crivellin}
\email{andreas.crivellin@cern.ch}
\affiliation{Paul Scherrer Institut, CH--5232 Villigen PSI, Switzerland}
\affiliation{Physik-Institut, Universit\"at Z\"urich, Winterthurerstrasse 190, CH--8057 Z\"urich, Switzerland}

\author{Francesco Saturnino}
\email{saturnino@itp.unibe.ch}
\affiliation{Albert Einstein Center for Fundamental Physics, Institute
	for Theoretical Physics,\\ University of Bern, CH-3012 Bern,
	Switzerland}

\begin{abstract}
In this article we investigate the correlations between tauonic $B$ meson decays (e.g. $B\to \tau\nu$, $B\to D^{(*)}\tau\nu$, $B\to \pi\tau\nu$) and electric dipole moments (EDMs), in particular the one of the neutron, in the context of the $S_1$ scalar leptoquark. We perform the matching of this model on the effective field theory taking into account the leading renormalization group effect for the relevant observables. We find that one can explain the hints for new physics in $b\to c\tau\nu$ transitions without violating bounds from other observables. Even more interesting, it can also give sizable effects in $B\to\tau\nu$, to be tested at BELLE II, which are correlated to (chromo) electric dipole operators receiving $m_\tau/m_u$ enhanced contributions. Therefore, given a deviation from the Standard Model (SM) expectations in $B\to\tau\nu$, this model predicts a sizable neutron EDM. In fact, even if new physics has CP conserving real couplings, the CKM matrix induces a complex phase and already a 10\% change of the $B\to\tau\nu$ branching ratio (with respect to the SM) will lead to an effect observable with the n2EDM experiment at PSI.
\end{abstract}


\maketitle

\section{Introduction}

In the past four decades, the Standard Model (SM) of particle physics has been extensively tested and its predictions were very successfully confirmed, both in high energy searches as well as in low energy precision experiments. However, it is well known that the SM cannot be the ultimate theory describing the fundamental constituents of matter and their interactions. For example, it cannot accommodate for the observed matter--antimatter asymmetry in the universe: For satisfying the Sakharov conditions~\cite{Sakharov:1967dj} the amount of CP violation within the SM is far too small~\cite{Cohen:1993nk,Gavela:1993ts,Huet:1994jb,Gavela:1994ds,Gavela:1994dt,Riotto:1999yt}. Therefore, additional sources of CP violation are required and such models in general lead to nonvanishing electric dipole moments of neutral fermions. Thus, EDMs are very promising places to search for physics beyond the SM (see e.g. Ref.~\cite{Chupp:2017rkp,Yamanaka:2017mef} for a recent review). However, the effect of new physics (NP) in EDMs decouples with the NP scale which is a priori unknown, unless new particles, or at least deviations from the SM in other precision observables, are found.

In this respect, tauonic $B$ decays are very promising channels for the (indirect) search for NP, especially in the light of the observed tensions between the SM predictions and experiments above the 3$\,\sigma$ level~\cite{Amhis:2016xyh}. These decays involve both down-type quarks and charged leptons of the third generation (i.e. bottom quarks and tau leptons) which are, due to their mass, very special and distinct from the fermions of the first two generations.\footnote{In group theory language, the SM possesses a global $U(3)^5$ flavor symmetry which is broken by the thrid generation Yukawa couplings to $U(2)^5$~\cite{Barbieri:1995uv}.} In fact, to explain these anomalies, TeV scale NP with order one couplings to the third generation is required. Note that the tensions in $b\to c\tau\nu$ transitions are supported by $b\to u\tau\nu$ data (i.e. $B \to \pi \tau \nu$ and $B \to \tau \nu$) and the forthcoming measurements of both $b\to c\tau\nu$  and $b\to u\tau\nu$ processes by LHCb and BELLE II will be able to confirm (or disprove) the presence of NP in these decays. 

Therefore, it is very interesting to investigate the possible impact of models which can give sizable effects in tauonic $B$ decays and EDMs. In this paper we choose the scalar leptoquark $S_1$ $SU(2)_L$ singlet which couples to SM fermions via the Lagrangian
\begin{align}
\mathcal{L} =& \left( \lambda_{fi}^{L}\overline {Q_f^c} i{\tau _2}{L_i}  +{\lambda}^{R}_{fi}\overline{u^c_f}\ell_i\right)\Phi_1^{\dagger} + {\rm{h}}{\rm{.c}}.\,.\label{LLQ}
\end{align}
Here,  $L$ ($Q^c$) is the lepton (charge conjugated quark) $SU(2)_L$ doublet,  $\ell$ ($u^c$) the charged lepton (charge conjugated up quark) singlet and $f,i$ are flavor indices. This model is theoretically well motivated since $S_1$ is present within the $R$-parity violating MSSM in the form of right-handed down squarks~\cite{Hall:1983id,Ross:1984yg,Barger:1989rk,Dreiner:1997uz,Barbier:2004ez}.\footnote{Note that in the minimal R-parity violating MSSM the coupling to charged conjugated fields in \eq{LLQ} is absent. For an analysis of EDM constraints within this setup see Ref.~\cite{Yamanaka:2014nba}.}

This leptoquark (LQ) is a prime candidate for providing the desired correlations between tauonic $B$ decays and EDMs. It possesses couplings to left- and right-handed quarks which is a necessary requirement for generating EDMs at the one-loop level~\cite{Fuyuto:2018scm,Dekens:2018bci}. It also contributes to $b\to c\tau\nu$ at tree level~\cite{Fajfer:2012jt,Deshpande:2012rr,Tanaka:2012nw,Sakaki:2013bfa,Freytsis:2015qca,Hati:2015awg,Bauer:2015knc,Li:2016vvp,Zhu:2016xdg,Popov:2016fzr,Deshpand:2016cpw,Crivellin:2017zlb,Altmannshofer:2017poe,Kamali:2018fhr,Azatov:2018knx,Wei:2018vmk,Hu:2018lmk,Angelescu:2018tyl,Kim:2018oih,Yan:2019hpm} and gives a very good fit to data (including polarization observables)~\cite{Feruglio:2018fxo,Iguro:2018vqb,Blanke:2018yud,Murgui:2019czp} since it generates vector, scalar and tensor operators. Similarly, it contributes to $b\to u\tau\nu$ transitions, in particular to $B\to\tau\nu$, where the situation becomes especially interesting. As we will see, in this case the model leads to $m_\tau/m_u$ enhanced CP violating effects in (chromo) electric dipole operators (see Fig.~\ref{Fig:EDM}) which are even present for real NP parameters due to the large phase contained in the CKM element $V_{ub}$.

This paper is structured as follows: In the next section we will calculate the contributions to the relevant observables and discuss their experimental status. Section~\ref{pheno} presents our phenomenological analysis before we conclude in Sec.~\ref{conclusions}.

\section{Observables and Contributions}
\label{observables}

In this section we discuss our setup, calculate the predictions for the relevant observables and discuss their current experimental situation and future prospects. 

After electroweak symmetry breaking, the Lagrangian in \eq{LLQ} decomposes into components 
\[{\cal L}_{\rm eff}^{\rm EW} \!=\! \left( {\lambda _{fi}^R\bar u_f^c{P_R}{\ell _i}\! +\! V_{fj}^*\lambda _{ji}^L\bar u_f^c{P_L}{\ell _i}\! -\! \lambda _{fi}^L\bar d_f^c{P_L}{\nu _i}} \right)\!\Phi _1^\dag  + {\rm{h}}.{\rm{c}}.\]
Here, we work in the down basis, meaning that the CKM matrix $V$ appears in the couplings to left-handed up-type quarks. We denote the mass of the LQ by $M$ and neglect its couplings to the SM Higgs boson which have a negligible phenomenological impact. The most relevant classes of observables in our model are $b\to s\nu\nu$ and $b\to c(u)\tau\nu$ transitions as well as EDMs, $D^0-\bar D^0$ mixing and $Z$-$\tau\tau$ as well as $W$-$\tau\nu$ couplings which we consider now in more detail.

\begin{boldmath}
\subsection{$b\to s\nu \nu$}
\end{boldmath}

For $b\to s\nu \nu$ transitions we follow the conventions of Ref.~\cite{Buras:2014fpa}
\begin{align}
		\mathcal{H}_{{\rm eff}}^{\nu\nu}&=-\dfrac{4G_F}{\sqrt{2}}V_{td_k}V_{td_j}^{*} \left(C^{fi}_{L,jk}\mathcal{O}_{L,jk}^{fi}+C^{fi}_{R,jk}\mathcal{O}_{R,jk}^{fi}\right)\,,
		\nonumber\\
		\mathcal{O}_{L(R),jk}^{fi}&=\frac{\alpha}{4\pi}\left[\bar{d}_{j} \gamma^{\mu}P_{L(R)}d_{k}\right] \left[\bar{\nu}_{f}\gamma_{\mu}\left(1-\gamma_5\right)\nu_i\right]\,,
\end{align}
and obtain, already at tree level, the contribution
\begin{align}
C_{L,jk}^{fi\,{\rm NP}} = \frac{{\sqrt 2 }}{{4{G_F}{V_{t{d_k}}}V_{t{d_j}}^*}}\frac{\pi }{\alpha }\frac{{\lambda _{jf}^{L*}\lambda _{ki}^L}}{{M_{}^2}}\,.\label{bsnunu}
\end{align}
Here the most relevant decays are $B\to K^{(*)}\nu\nu$ for which $C_{L,sb}^{{\rm SM},fi}\approx-1.47/s_W^2\delta_{fi}$ and branching ratios, normalized by the corresponding SM predictions, read
\begin{equation}
{R_{K^{(*)}}^{\nu\bar{\nu}}} = 
\frac{1}{3}\sum\limits_{f,i=1}^3 \dfrac{ \big|{C_{L,sb}^{fi}}\big|^2}{\big|{C_{L,sb}^{{\rm SM},ii}}\big|^2} \,.
\end{equation}
This has to be compared to the current experimental limits ${R_K^{\nu\bar{\nu}}} < 3.9$ and ${R_{{K^*}}^{\nu\bar{\nu}}} < 2.7$~\cite{Grygier:2017tzo} (both at $90\%\,\mathrm{C.L.}$). The future BELLE II sensitivity for $B\to K^{(*)}\nu\bar{\nu}$ is 30\% of the SM branching ratio~\cite{Abe:2010gxa}.
\medskip

\begin{boldmath}
\subsection{$b\to c(u)\tau\nu$}
\end{boldmath}

For tauonic $B$ decays we define the effective Hamiltonian as
\begin{align}
{\cal H}_{{\rm{eff}}}^{\tau \nu } = \frac{{4{G_F}}}{{\sqrt 2 }}{V_{{u_f}b}}\left( {C_{VL}^fO_{VL}^f + C_{SL}^fO_{SL}^f 
	+ C_{TL}^fO_{TL}^f} \right)\,,\nonumber
\end{align}
with the operators given by
\begin{align}
\begin{aligned}
	O_{VL}^{u_f} &= {{\bar u}_f}{\gamma ^\mu }{P_L}b\bar \tau {\gamma _\mu }{P_L}{\nu _\tau }\,,\\
	O_{SL}^{u_f} &= {{\bar u}_f}{P_L}b\bar \tau {P_L}{\nu _\tau }\,,\\
	O_{TL}^{u_f} &= {{\bar u}_f}{\sigma ^{\mu \nu }}{P_L}b\bar \tau {\sigma _{\mu \nu }}P_L{\nu _\tau }\,.
\end{aligned}
\end{align} 
In the SM $C_{VL}^{u_f}=1$ and our NP matching contributions at tree level are given by
\begin{align}
		\begin{aligned}
C_{VL}^{u_f} &= \frac{{\sqrt 2 }}{{8{G_F} \, V_{{u_f}b}}}\frac{V_{{u_f}i}{\lambda _{i3}^{L*}\lambda _{33}^L}}{{M_{}^2}}\,, \\
C_{SL}^{u_f} &=  - 4C_{TL}^{u_f} = \frac{{ - \sqrt 2 }}{{8{G_F}V_{u_fb}}}\frac{{\lambda _{f3}^{R*}\lambda _{33}^L}}{{M_{}^2}}\,.
	\end{aligned}
\end{align}
Taking into account the QCD effects of Ref.~\cite{Aebischer:2018acj} to the matching, the one-loop EW and two-loop QCD renormalization group equation (RGE) for the scalar and tensor operators~\cite{Gracey:2000am,Gonzalez-Alonso:2017iyc} can be taken consistently into account. Numerically, this RGE evolution is given by 
\begin{align}
\left( \begin{array}{c}
C_{SL}^{u_f}(m_b) \\ C_T^{u_f}(m_b)           
\end{array}\right) &\approx  \left( \begin{array}{rr}
1.75 & -0.29 \\
0 & 0.84 
\end{array}\right)
\left( \begin{array}{c}
C_{SL}^{u_f}(1\,\mbox{TeV}) \\ C_{T}^{u_f}(1\,\mbox{TeV})    
\end{array}\right) , \nonumber
\end{align}
for a matching scale of $1\,\rm{TeV}$. Finally, the, ratios $R(D^{(*)})= \frac{\text{Br}[B \to D^{(*)} \tau \nu]}{ \text{Br}[B \to D^{(*)} \ell \nu]}$ with $\ell=\{\mu,e\}$ in terms of the Wilson coefficients at the $b$ scale are given by~\cite{Blanke:2018yud}
\begin{align}
\begin{aligned}
&\frac{{R(D)}}{{{R_{{\rm{SM}}}}(D)}} \simeq |1 + C_{VL}^c{|^2} + 1.54{\mkern 1mu} \Re [(1 + C_{VL}^c)C_{SL}^{c * }]  \\
& + 1.09|C_{SL}^c{|^2}+ 1.04{\mkern 1mu} \Re [(1 + C_{VL}^c)C_T^{c * }] + 0.75|C_T^c{|^2}{\mkern 1mu} ,\\ 
&\frac{{R({D^*})}}{{{R_{{\rm{SM}}}}({D^*})}} \simeq |1 + C_{VL}^c{|^2} - 0.13{\mkern 1mu} \Re [(1 + C_{VL}^c)C_{SL}^{c * }] \\ & + 0.05|C_{SL}^c{|^2} - 5.0{\mkern 1mu} \Re [(1 + C_{VL}^c)C_T^{c * }]+16.27|C_T^c{|^2}\,.
\end{aligned}
\end{align}
Similarly, for $b\to u\tau\nu$ transitions we have
\begin{equation}
\frac{{{\rm Br}\left[ {B \to \tau \nu } \right]}}{{{\rm Br}{{\left[ {B \to \tau \nu } \right]}_{SM}}}} = {\left| {1 + C_{VL}^u - \frac{{m_B^2C_{SL}^u}}{{{m_b}{m_\tau }}}} \right|^2}\,.
\end{equation}
The corresponding formula for $B\to\pi\tau\nu$ can be found in Ref.~\cite{Tanaka:2016ijq}. However, here the effect of scalar and tensor operators is much smaller, making the theoretically very clean $B \to \tau \nu$ decays the primary place to search for them.

Combining the experimental measurements of $b\to c\tau\nu$ transitions from LHCb~\cite{Aaij:2015yra,Aaij:2017uff,Aaij:2017deq}, Belle~\cite{Huschle:2015rga,Sato:2016svk,Hirose:2016wfn,Hirose:2017dxl,Abdesselam:2019dgh} and Babar~\cite{Lees:2012xj,Lees:2013uzd}, one finds a combined tension of $3.1\,\sigma$ in $R(D^{(*)})$~\cite{Amhis:2016xyh}\footnote{In Ref.~\cite{Ikeno:2019tkh} it was shown that uncertainties from meson exchanges between initial and final states might be bigger than the estimated SM uncertainty, which could alleviate the tension in $R(D^{(*)})$. On the other hand, recent improvements in form factor calculations~\cite{Bordone:2019vic} lower the SM prediction and increase the tension. These two effects are not included in Ref.~\cite{Amhis:2016xyh} but will not change the result significantly.}. However, note that here the $B_c\to J/\Psi\tau\nu$ measurement of LHCb~\cite{Aaij:2017tyk}, which also lies significantly above the SM prediction, is not included.\footnote{See Ref.~\cite{Watanabe:2017mip,Chauhan:2017uil} for an analysis including $B_c\to J/\Psi\tau\nu$ before the latest BELLE update~\cite{Abdesselam:2019dgh}.} In $b\to u\tau\nu$ transitions, the theory prediction for $B\to\tau\nu$ crucially depends on $V_{ub}$. While previous lattice calculations resulted in rather small values of $V_{ub}$, recent calculations give a larger value (see Ref.~\cite{Ricciardi:2016pmh} for an overview). However, the measurement is still above the SM prediction by more than 1$\,\sigma$, as can be seen from the global fit~\cite{Charles:2004jd}. In $R(\pi)=\frac{{\rm Br}[B\to\pi\tau\nu]}{{\rm Br}[B\to\pi\ell\nu]}$ there is also a small disagreement between theory~\cite{Bernlochner:2015mya} and experiment~\cite{Hamer:2015jsa} which does not depend on $V_{ub}$, once more pointing towards an enhancement. Therefore, even though the $b\to u\tau\nu$ results are not significant on their own, they point in the same direction as $b\to c\tau\nu$ (i.e. towards an enhancement with respect to the SM) and thus strengthen the case for NP in tauonic $B$ decays.
\medskip

\subsection{EDMs}

\begin{figure}[t]
	\begin{center}
		\begin{tabular}{cp{7mm}c}
			\includegraphics[width=0.32\textwidth]{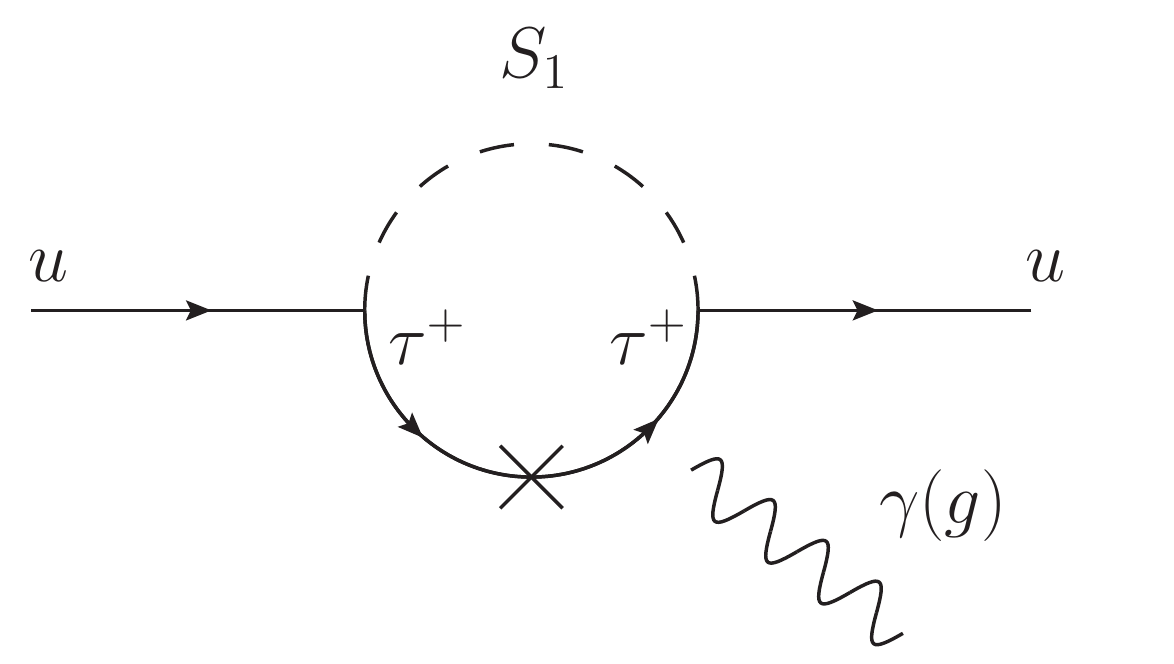}
		\end{tabular}
	\end{center}
	\caption{Feynman diagram showing the contribution of our model to the dipole operators of \eq{EDMoperators}. The cross denotes the chirality flip by the tau mass which leads to the crucial $m_\tau/m_u$ enhancement.}         
	\label{Fig:EDM}
\end{figure}

For EDMs the relevant Hamiltonian in our case is
\begin{equation}
{\cal H}_{{\rm{eff}}}^{{\rm{nEDM}}} = C_\gamma ^uO_\gamma ^u + C_g^uO_g^u 
+ C_T^{u\tau }O_T^{u\tau }\,,
\end{equation}
with 
\begin{align}
\begin{aligned}
O_\gamma ^u &= e\bar u{\sigma^{\mu \nu }}{P_R}u{F_{\mu \nu }}\,,\\
O_g^u &= {g_s}\bar u{\sigma^{\mu \nu }}{P_R}u \, T^a{G_{\mu \nu }^a}\,,\\
O_T^{u\tau } &= \bar u{\sigma _{\mu \nu }}{P_R}u\bar \tau {\sigma ^{\mu \nu }}{P_R}\tau \,.
\end{aligned}
\label{EDMoperators}
\end{align}
At the high scale we find the matching contributions (depicted in Fig.~\ref{Fig:EDM})
\begin{align}
\begin{aligned}
C_T^{u\tau } &= -\frac{{V_{1j}^{}\lambda _{j3}^{L*}}\lambda _{13}^R}{{8{M^2}}}\,,\\
C_\gamma ^u &=  - \frac{{{m_\tau }{V_{ub}}}}{{96{\pi ^2}{M^2}}}\lambda _{33}^{L*}\lambda _{13}^R\left( {4 + 3\log \left( {{\mu ^2}/{M^2}} \right)} \right)\,,\\
C_g^u &=  - \frac{{{m_\tau }{V_{ub}}}}{{64{\pi ^2}{M^2}}}\lambda _{33}^{L*}\lambda _{13}^R\,.
\end{aligned}
\label{EDMmatching}
\end{align}
Note that we only get up-quark contributions since we do not have (at the one-loop level) CP violating couplings to down-type quarks. Importantly, note that our effect in $C_\gamma ^u$ and $C_g^u$ is parametrically enhanced by $m_\tau/m_u$, making a sizable effect in EDMs possible. This enhancement of the dipole operators also allows us to safely neglect the effects of charm quarks, four-fermion operators and of the Weinberg operator otherwise relevant for LQs~\cite{Dekens:2018bci}.

Next, we use the one-loop RGE to evolve these Wilson coefficients of \eq{EDMmatching} down to the neutron scale. Here, combining and adjusting the results of Ref.~\cite{Crivellin:2017rmk} and Ref.~\cite{Borzumati:1998tg} to our case we obtain\footnote{For the same RGE in a different operator basis see Ref.~\cite{Hisano:2012cc}.}
\[\mu \frac{d}{{d\mu }}\!\left( {\begin{array}{*{20}{c}}
	{C_T^{u\tau }}\\
	{C_\gamma ^u}\\
	{C_g^u}
	\end{array}} \right) \!=\! \left( {\begin{array}{*{20}{c}}
	{\frac{{{C_F}{\alpha _s}}}{{2\pi }}}&0&0\\
	{ - \frac{{{m_\tau }}}{{2\pi^2}}}&{\frac{{{\alpha _s}{C_F}}}{2\pi }}&{\frac{{4{C_F}{\alpha _s}}}{{3\pi }}}\\
	0&0&{\frac{{{\alpha _s\left( {10{C_F} - 12} \right)}}}{{4\pi }}}
	\end{array}} \right)\!\!\left( {\begin{array}{*{20}{c}}
	{C_T^{u\tau }}\\
	{C_\gamma ^u}\\
	{C_g^u}
	\end{array}} \right).\]
The solution to this differential equation can be written in terms of an evolution matrix in the form 
\begin{equation}
\vec C\left( {{\mu _l}} \right) = U\left( {{\mu _l},{\mu _h}} \right)\vec C\left( {{\mu _h}} \right)
\end{equation}
with
\begin{align}
U\left( {{\mu _l},{\mu _h}} \right) &\!=\! \left(\!\!\! {\begin{array}{*{20}{c}}
	{{\eta ^{\frac{4 }{{3{\beta _0}}}}}}&0&0\\
	{ - {{\rm{m}}_\tau }X}&{{\eta ^{\frac{{4}}{{3{\beta _0}}}}}}&{\frac{{16}}{3}{\eta ^{\frac{{14}}{{3{\beta _0}}}}}\left( {{\eta ^{\frac{2}{{3{\beta _0}}}}} - 1} \right)}\\
	0&0&{{\eta ^{\frac{{2}}{{3{\beta _0}}}}}}
	\end{array}} \!\!\!\right),\\[8pt]
{\beta _0} &= \frac{{33 - 2f}}{3},\;\;\eta  = \frac{{{\alpha _s}\left( {{\mu _h}} \right)}}{{{\alpha _s}\left( {{\mu _l}} \right)}}\,,\intertext{and}
X &= \frac{{{\eta ^{\frac{{\rm{4}}}{{3{\beta _0}}}}}\left( {{\eta ^{\frac{{\rm{4}}}{{{\beta _0}}}}} - 1} \right){\beta _0}}}{{8\pi^2 \, \log(\eta) }}{\rm log}\left(\frac{\mu_l}{\mu_h}\right)\,,
\end{align}
where $f$ is the number of active quark flavors. The final evolution matrix is obtained by running with the appropriate numbers of flavours from the LQ scale down to 1 GeV. 

Finally, the effects in the neutron and proton EDMs are given by~\cite{Cirigliano:2016nyn} 
\begin{align}
{d_n}/e &= -\left( {0.44 \pm 0.06} \right){\mathop{\rm Im}\nolimits} \left[ {C_\gamma ^u} \right] - \left( {1.10 \pm 0.56} \right){\mathop{\rm Im}\nolimits} \left[ {C_g^u} \right],\nonumber\\
{d_p}/e &= \left( {1.48 \pm 0.14} \right){\mathop{\rm Im}\nolimits} \left[ {C_\gamma ^u} \right] + \left( {2.6 \pm 1.3} \right){\mathop{\rm Im}\nolimits} \left[ {C_g^u} \right]\,,\nonumber
\end{align}
in terms of the Wilson coefficients evaluated at 1 GeV. The neutron and proton EDMs then enter atomic ones, most importantly in mercury and deuteron (see Ref.~\cite{Cirigliano:2016nyn} for details). 

On the experimental side, $d_{Hg}$~\cite{Graner:2016ses} gives currently slightly better bounds than the neutron EDM, while the one of the proton and the deuteron is not measured yet. However, $d_p$ and $d_D$ will be very precisely known from future experiments~\cite{Pretz:2013us,Eversmann:2015jnk} and concerning $d_n$ there will be soon an improvement of one order of magnitude in sensitivity compared to the current limit of $3.6\times 10^{-26}e\,cm$~\cite{Baker:2006ts,Afach:2015sja} from the n2EDM experiment at PSI~\cite{Abel:2018yeo}. Therefore, we will focus on $d_n$ in our phenomenological analysis.

\begin{figure*}[t]
	\begin{center}
		\begin{tabular}{cp{7mm}c}
\includegraphics[width=\textwidth]{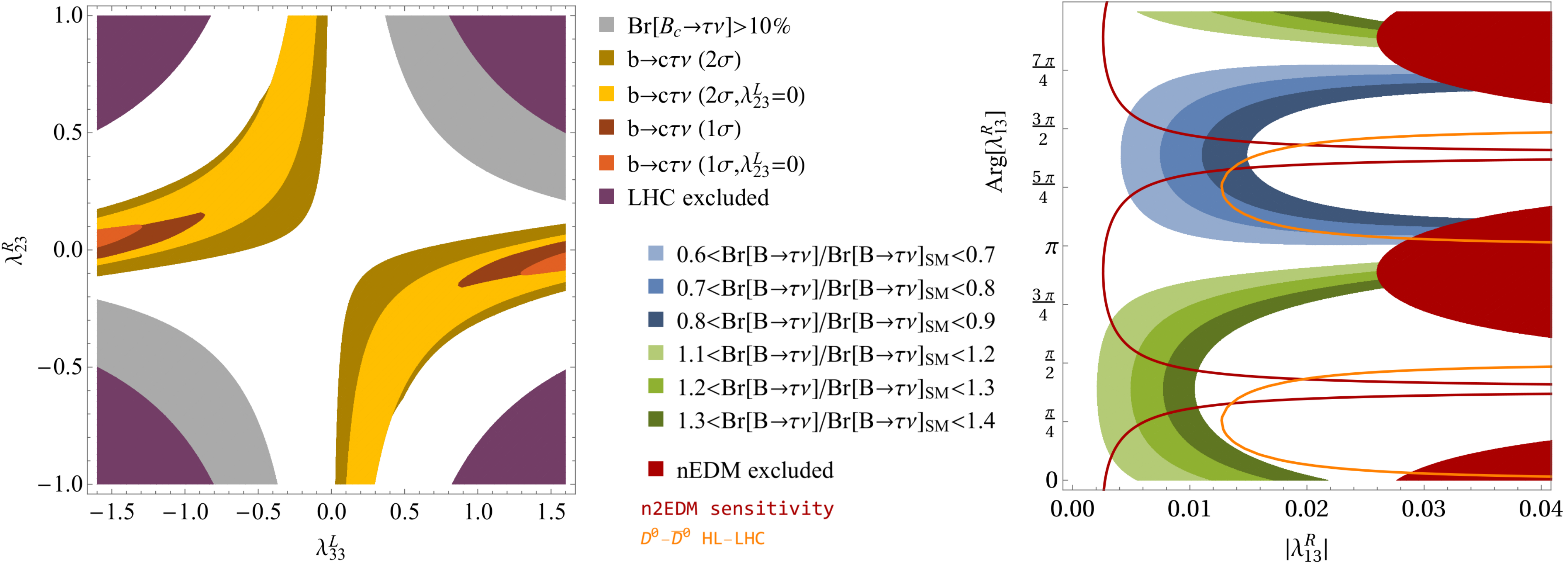}
		\end{tabular}
	\end{center}
	\caption{Left: preferred regions in the $\lambda_{33}^L$--$\lambda_{23}^R$ plane from $b\to c\tau\nu$ data for $M=1\,$TeV. Here, both the case of $\lambda_{23}^L=0$ and the one taking the maximally allowed value of $\lambda_{23}^L$ from $B\to K^*\nu\nu$ are shown. A good fit to data requires $|\lambda_{33}^L|\approx 1$ in both cases. Note that our model is compatible with LHC searches for monotaus and with $B_c$ lifetime constraints which exclude the dark pink and gray regions. Right: The green (blue) regions indicate where $B\to\tau\nu$ is enhanced (suppressed) by 10\%-40\% w.r.t. the SM for $M=1\,$TeV, $\lambda_{33}^L=1$ and $\lambda_{13}^L=0$.	The dark red region is excluded by the neutron EDM and the dark red contour denotes the n2EDM sensitivity. The orange contour shows the HL-LHC sensitivity to CP violation in $D_0-\bar D_0$ mixing which is nicely complementary to EDM searches.}         
	\label{Fig:bctaunu}
\end{figure*}

\begin{boldmath}
\subsection{$D_0-\bar D_0$ mixing}
\end{boldmath}
To describe $D_0-\bar D_0$ mixing we use the effective Hamiltonian
\begin{align*}
\mathcal{H}^{D\bar{D}}_{\rm{eff}}&=C_{1}^\prime Q_1^\prime\,, \;\; Q_1^\prime=\left[\bar{u}_{\alpha}\gamma_{\mu}P_{R}c_{\alpha}\right]\left[\bar{u}_{\beta}\gamma^{\mu}P_{R}c_{\beta}\right]\,,
\end{align*}
and find at the high scale
\begin{equation}
C_1^\prime=\dfrac{\left(\lambda_{13}^R\lambda_{23}^{R*}\right)^2}{128\pi^2 M^2}\,,
\end{equation}
from the one-loop matching. The evolution of $C_1^\prime$ was calculated in Refs.~\cite{Buras:2000if,Ciuchini:1997bw} and yields approximately~\cite{Golowich:2007ka}
\begin{align}
C_1^\prime(3 \, \text{GeV}) \approx 0.8 C_1^\prime (1 \, \text{TeV})\,.
\end{align}
The matrix element for the $D$-meson mixing is given by
\begin{align}
\langle \bar{D}^0|Q_1^\prime(\mu)|D^0\rangle=\frac{1}{3} B_1(\mu) m_D f_D^2 \,,
\end{align}
where $B_1(\mu)=0.75$ at the scale $\mu=3\,\rm{GeV}$ \cite{Carrasco:2014uya}. The mass difference in the $D$-meson system is given by
\begin{align}
\Delta m_D = 2 {\text{Re}\left[\langle\bar{D}^0|\mathcal{H}^{D\bar{D}}_{\rm{eff}}|D^0\rangle\right]}  { \equiv 2 \text{Re}\left[M_{12}\right]} \ .
\end{align}
Further, we write
\begin{align}
\sin \phi_{12} =  -\frac{2\text{Im}\left[ M_{12}\right]}{\Delta m_D}  \ .
\end{align}
The averages of the experimental values read~\cite{Bevan:2014tha,Bazavov:2017lyh}
\begin{align}
	\begin{aligned}
0.001&<|M_{12}|[{\rm{ps}^{-1}}] < 0.008 \ , \\
-3.5&<\phi_{12}[{}^\circ] < 3.3  \ , \\
f_D &= 212 \, \rm{MeV} \ ,
	\end{aligned}
\end{align}
at 95\% C.L. At a high luminosity LHC (HL-LHC) the sensitivity to $\phi_{12}$ could be improved down to the SM expectation of $\approx 0.17^\circ$~\cite{Cerri:2018ypt}.
\medskip

\begin{boldmath}
\subsection{$W \to \tau \nu$ and $Z \to \tau \tau$}
\end{boldmath}
Virtual corrections with top quarks and LQs modify couplings of gauge bosons to charged leptons, in particular to the tau.  Parametrizing the interactions as
\begin{align*}
-\mathcal{L}=\frac{g_2}{\sqrt{2}}\Lambda_{3 i}^{W} \left(\bar{\tau}\gamma^{\mu}P_{L}\nu_{i}W_{\mu}^{-}\right) +\frac{g_2}{2 c_w}\bar{\tau} \gamma^{\mu} \left( \Lambda^V   -\Lambda^A \gamma_5\right) \tau Z_{\mu}
\end{align*}
with
\begin{align*}
\Lambda_{3i}^{W}&=\delta_{3i}+\Lambda_{3i}^{\rm{LQ}} \ ,\;
	\Lambda^{V,A}=\Lambda^{V,A}_{\text{SM}} + \Delta^{V,A}_{\rm LQ} \ , \\ \Lambda^V_{\rm{SM}}&=-\frac{1}{2}+2s_w^2 \ , \ \Lambda^{A}_{\rm{SM}}=-\frac{1}{2} \ ,
\end{align*}
the LQ effects at $q^2=0$ (the contributions proportional to  gauge boson mass are suppressed) are given by
\begin{align}
\Lambda_{3i}^{LQ}=&\frac{N_c   m_t^2 }{192 \pi ^2 M^2}\Bigg[3V_{3h} \lambda _{h3}^{L*} V_{3k}^*\lambda _{ki}^{L} \left(1+2 \log \left(\frac{m_t^2}{M^2}\right)\right) \Bigg] \ ,\nonumber\\
\Delta^L_{\rm LQ}&=V_{3l}\lambda_{l3}^{L*}V_{3a}^{*}\lambda_{a3}^{L}
\frac{N_c \, m_t^2}{32 \pi^2 \, M^2} \left[ 1 + \log \left(\frac{m_t^2}{M^2} \right) \right] \ , \nonumber\\
\Delta^R_{\rm LQ}&= -\lambda^{R*}_{33}\, \lambda^{R}_{33} \frac{ N_c  \, {m_t^2}}{32 \pi^2 \, M^2} \left[ 1 + \log \left(\frac{m_t^2}{M^2} \right) \right] \ ,
\end{align}
with $
	\Delta^{V}_{\rm LQ}=-\Delta^L_{\rm LQ} - \Delta^ R_{\rm LQ}$ and $	\Delta^A_{\rm LQ} = \Delta^R_{\rm LQ} - \Delta^L_{\rm LQ}$.
This leads to $
{|\Lambda_{33}^W|}=\left|1+\Lambda_{33}^{\rm{LQ}}\right|$.
Experimentally, the averaged modification of the $W$-$\tau \nu$ coupling extracted from $\tau \to \mu \nu \nu$ and $\tau \to e \nu \nu$ decays reads (averaging the central value but with unchanged error) \cite{Pich:2013lsa,Tanabashi:2018oca}
\begin{align}
{|\Lambda_{33}^{W \, \rm{exp}}|}\approx 1.002 \pm 0.0015 \ ,
\end{align}
which provides a better constraint than data of $W$ decays.

Concerning $Z\to\tau\tau$ the axial vector coupling is much better constrained that the vectorial one\cite{Pich:2013lsa,Tanabashi:2018oca}
\begin{align}
{\Lambda^A_{\rm{exp}}}/{\Lambda^A_{\rm{SM}}} = 1.0019 \pm 0.0015 \ ,
\end{align}
with
${\Lambda^A}/{\Lambda^A_{\rm{SM}}}=1+2 \Delta^L_{\rm LQ}-2 \Delta^R_{\rm LQ}$.
\medskip

\section{Phenomenology}
\label{pheno}

Looking at the phenomenological consequences of our model, note that couplings to muons or electrons are obviously not necessary to obtain the desired effects in tauonic $B$ decays. Even though our $S_1$ model can in principle account for the anomalous magnetic moment of the muon~\cite{Djouadi:1989md,Davidson:1993qk,Couture:1995he,Chakraverty:2001yg,Cheung:2001ip,Mahanta:2001yc,Queiroz:2014pra,Bauer:2015knc,Das:2016vkr,Biggio:2016wyy,Crivellin:2018qmi} (or electron~\cite{Crivellin:2018qmi}) via a $m_t/m_\mu$ enhanced effect, this is not possible in the presence of large couplings to tau leptons since also here $m_t$ enhanced effects generate too large rates of $\tau\to\mu(e)\gamma$. Similarly, our model cannot address the $b\to s\mu^+\mu^-$ anomalies if one aims at a sizable effect in tauonic $B$ decays~\cite{Becirevic:2016oho}. Therefore, we will disregard (i.e. set to zero) the couplings to muons and electrons. Couplings to top-quarks affect $\tau\to\mu\nu\nu$~\cite{Feruglio:2016gvd} and $Z\to\tau^+\tau^-$~\cite{Arnan:2019olv}. Here we see that
$\Delta^L\approx -0.0006 |\lambda_{33}^L|^2$ and $\Lambda_{33}^{\rm{LQ}} \approx -0.0008 |\lambda_{33}^L|^2$ (for $M=1\,$TeV) is compatible with experiments for $|\lambda_{33}^L|  < 1$. Note that we improve the agreement in $Z\to\tau\tau$ data while slightly worsening $\tau\to \ell\nu\nu$ data, which is already a bit away from the SM prediction.

Thus, we are left with $\lambda _{13}^R$,$\lambda _{23}^R$ and $\lambda _{i3}^L$ as free parameters for studying the effect in tauonic $B$ decays and the correlations with EDMs. In the following we will set $M=1\,$TeV which is also well compatible with the latest direct search results of CMS for third generation LQs~\cite{Takahashi:2019zsl,Sirunyan:2018ruf}.\footnote{More sophisticated analysis of LHC data can be found in Refs.~\cite{Raj:2016aky,Bansal:2018eha,Mandal:2018kau}. However, since for t-channel exchange the EFT limits are in general stronger than the ones in the UV complete model, we will use for simplicity the results of Ref.~\cite{Greljo:2018tzh} in the following which show that 1 TeV is compatible with data.} 

Let us now turn to $b\to c\tau\nu$ processes, where effects of the order of 10\% compared to the corresponding tree-level SM amplitude are required. Since our model can give (according to \eq{bsnunu}) tree-level effects in $B\to K^{(*)}\nu\nu$ decays (which are loop suppressed in the SM), these  contributions must be suppressed. Since the bottom coupling to taus should be sizable, the coupling to strange quarks is tightly bound. We show the preferred regions, according to the updated global fit of Ref.~\cite{Blanke:2018yud}, from $b\to c\tau\nu$ processes in the left plot of Fig.~\ref{Fig:bctaunu}. These regions are shown for $\lambda^L_{23}=0$ but also the possible impact of $\lambda^L_{23}\neq0$, taking its maximally allowed values from $B\to K^*\nu\nu$, is depicted. Note that our model is not in conflict with the $B_c$ lifetime~\cite{Celis:2016azn,Alonso:2016oyd} (in fact, it is even compatible with the 10\% limit of Ref.~\cite{Akeroyd:2017mhr}) nor with direct LHC searches for monotaus~\cite{Greljo:2018tzh}. So far we worked with real parameters in order to maximize the effect in $R(D^{(*)})$. However, even for complex couplings the effect in nuclear and atomic EDMs would be strongly suppressed since only up and down quarks contribute directly to these observables. 

Therefore, let us now turn to $b\to u\tau\nu$ where couplings to up quarks are obviously needed. Here, even for real couplings an effect in the neutron EDM is generated due to the large phase of $V_{ub}$. This effect could only be avoided for ${\rm Arg}[\lambda _{13}^{R*}\lambda _{33}^L]={\rm Arg}[V_{ub}]$. However, since there is no (obvious) symmetry which could impose this relation, such a configuration would be fine-tuning. This can be seen from the right plot in Fig.~\ref{Fig:bctaunu}, where we show the predictions for ${\rm Br}[B\to\tau\nu]/{\rm Br}[B\to\tau\nu]_{\rm SM}$ as a function of the absolute value and the phase of $\lambda_{13}^R$ for $\lambda^{L}_{33}=1$ (as preferred by $b\to c\tau\nu$ data). The dark red contour lines denote the n2EDM sensitivity, showing that a 10\% effect in $B\to\tau\nu$ with respect to the SM will lead to an observable effect in the neutron EDM within our model. Finally, taking $\lambda^R_{23}=-0.1$, as preferred by $b\to c\tau\nu$ (see left plot of Fig.~\ref{Fig:bctaunu}), CP violation in $D^0-\bar D^0$ mixing is generated. Here the red contour denotes the future HL-LHC sensitivity which is complementary to the region covered by EDM searches.

\section{Conclusions}
\label{conclusions}

In this article we studied the interplay between tauonic $B$ meson decays and EDMs (in particular the one of the neutron) in a model with a scalar LQ $SU(2)_L$ singlet which can be identified with the right-handed down squark in the R-parity violating MSSM. We found that in order to explain the intriguing tensions in $b\to c\tau\nu$ data, $\lambda^L_{33}$ must be sizable and also a coupling to right-handed charm quarks and tau-leptons ($\lambda^R_{23}$) is required. In this setup, the model gives a very good fit to data and is compatible with $b\to s\nu\nu$ observables, LHC searches and $B_c$ lifetime constraints. Extending this analysis to  $b\to u\tau\nu$ transitions, in particular $B\to\tau\nu$, again right-handed couplings to up quarks ($\lambda^R_{13}$) are required to have a sizable effect. This leads to very important $m_\tau/m_u$ enhanced effects in (chromo) electric dipole operators generating in turn EDMs of nucleons and atoms. In particular, even for real couplings of the LQ to fermions, the large phase of $V_{ub}$ generates a sizable contribution to the neutron EDM. In fact, this effect should already be observable in the n2EDM experiment at PSI, assuming that, within our model, $B\to\tau\nu$ is enhanced (or suppressed) by around 10\% with respect to the SM.

\smallskip

{\it Acknowledgments} --- {The work of A.C. is supported by a Professorship Grant (PP00P2\_176884) of the Swiss National Science Foundation. The work of F.S. is supported by the Swiss National Foundation under Grant No. 200020\_175449/1. We thank Dario M\"uller for collaboration in the early stages of this article and Christoph Greub for useful comments on the manuscript. We are grateful to David Straub and Jason Aebischer for reminding us of the importance of $Z$-$\tau\tau$ and $W$-$\tau\nu$ couplings.}

\bibliography{BIB}

\end{document}